# HIGH-MASS, FOUR-PLANET CONFIGURATIONS FOR HR 8799: CONSTRAINING THE ORBITAL INCLINATION AND AGE OF THE SYSTEM

Jeffrey J. Sudol[1], Nader Haghighipour[2]

[1]West Chester University, 720 S. Church St., West Chester, PA, 19383, USA; jsudol@wcupa.edu.  [2]Institute for Astronomy and NASA Astrobiology Institute, University of Hawaii-Manoa, 2680 Woodlawn Drive, Honolulu, HI, 96822, USA; nader@ifa.hawaii.edu

# ABSTRACT

Debates regarding the age and inclination of the planetary system orbiting HR 8799, and the release of additional astrometric data following the discovery of the fourth planet prompted us to examine the possibility of constraining these two quantities by studying the long-term stability of this system at different orbital inclinations and in its high-mass configuration (7-10-10-10 $M_{Jup}$). We carried out ~1.5 million $N$-body integrations for different combinations of orbital elements of the four planets. The most dynamically stable combinations survived less than ~5 Myr at inclinations of 0° and 13°, and 41, 46, and 31 Myr at 18°, 23°, and 30°, respectively. Given such short lifetimes and the location of the system on the age-luminosity diagram for low-mass objects, the most reasonable conclusion of our study is that the planetary masses are less than 7-10-10-10 $M_{Jup}$ and the system is quite young. Two trends to note from our work are as follows. (1) In the most stable systems, the higher the inclination, the more the coordinates for planets b and c diverge from the oldest archival astrometric data (released after we completed our $N$-body integrations), suggesting that either these planets are in eccentric orbits or have lower orbital inclinations than that of planet d. (2) The most stable systems place planet e closer to the central star than is observed, supporting the conclusion that the planets are more massive and the system is young. We present the details of our simulations and discuss the implications of the results.

## 1. INTRODUCTION

The planetary system orbiting HR 8799 represents a unique astrophysical laboratory in that it is the only system for which direct images of multiple planets are available at the moment. The age of this system is critical to understanding how its planets formed and in calibrating the age-luminosity relationship for sub-stellar objects. However, the age of this system remains somewhat uncertain.

During the past decade, several efforts were made to estimate the age of HR 8799. Based on a comparison of $uvby\beta$ photometry to theoretical log $T_{eff}$ vs. log $g$ evolutionary tracks, Song et al. (2001) estimated an age of 50 to 1128 Myr, with the best value being 732 Myr. Chen et al. (2006), however, set the best value at 590 Myr. Later attempts pointed to a much younger system. For instance, Moór et al. (2006) estimated the age of the system to be between 20 and 150 Myr, based on the infrared excess of the debris disks. This result is consistent with the work of Rhee et al. (2007), who noted that the position of HR 8799 on a Hertzsprung-Russell diagram for A-class stars (cf. Lowrance et al. 2000) gives an age of 30 Myr. Marois et al. (2008) also examined the age of the system, and, based on several lines of evidence, concluded that it lies between 30 and 160 Myr. Recently, Doyon et al. (2010) and Zuckerman et al. (2011) independently identified HR 8799 as a likely member of the Columba Association, which is estimated to be 30 Myr old.

An important factor in constraining the age of the HR 8799 planetary system is its inclination. Based on asteroseismological modeling, Moya et. al. (2010) estimated that the age of the system lies between 26 and 430 Myr, or between 1123 and 1625 Myr, for an inclination of 36°. They also showed that another viable angle of inclination is 50°, in which case the age of the system lies between 1126 and 1486 Myr. Their methods and the data available at the time, however, did not permit them to investigate inclination angles between 18° and 36°. Using asteroseismological techniques and more extensive data, Wright et al. (2011) showed that the inclination angle must be greater than or equal to ~40°. Of course, asteroseismological techniques address the inclination of the star, not the planetary system, and the alignment of the two, although quite probable, is not a requirement.

The general consensus at this point is that the inclination of the planetary system orbiting HR 8799 is greater than 0° but less than 30°. Lafrenière et al. (2009) found that the best fits to the astrometric data for planet b are consistent with a low inclination orbit between 13° and 23°. Based on dynamical simulations of planets b, c, and d prior to the discovery of the fourth planet, Reidemeister et al. (2009) concluded that the inclination must be greater than 20° in order for the system to be stable. Based on the observations and modeling of the two debris disks orbiting HR 8799, Su et al. (2009) suggested that the inclination is less than 25°. Bergfors et al. (2011) showed that the astrometric data for planet d is inconsistent with an inclination angle of 0°. Recently, Soummer et al. (2011) added astrometric data from archival *Hubble Space Telescope* images to the mix and were able to rule out a number of orbital configurations. These authors also found that, if the planets are in coplanar orbits and reside in a 1d:2c:4b mean motion resonance (MMR), the astrometric data for planets b, c, and d favor an inclination of 28°.

The masses of the planets are another important factor in constraining the age of the HR 8799 planetary system. Planetary flux decreases as a planet cools, so, the older the planetary system, the more massive the planets must be in order to match the observed luminosities. Marois et al. (2008, 2010) deduced mass ranges of 5-11 $M_{Jup}$ for planet b and 7-13 $M_{Jup}$ for planets c, d, and e by comparing the luminosities of the planets to a theoretical age-luminosity diagram for low-mass objects and taking the age of the system to lie between 30 and 160 Myr. To further constrain the masses of the planets, Marois et al. (2010) investigated the stability of the planetary system. They considered two combinations of planetary masses: a low-mass configuration (5-7-7-7 $M_{Jup}$), corresponding to an age of 30 Myr, and a high-mass configuration (7-10-10-10 $M_{Jup}$), corresponding to an age of 60 Myr for standard, hot-start planet cooling models. They held the orbital elements of planets b, c, and d fixed to those values found by Fabrycky & Murray-Clay (2010) matching either (i) a 1:2 MMR between planets c and d, or (ii) a 1:2:4 MMR between planets b, c, and d. Marois et al. then carried out 100,000 numerical integrations of the orbits of the four planets, allowing the orbital elements for planet e to vary. The results of their integrations pointed to 12 low-mass and two high-mass systems with lifetimes in excess of 100 Myr. The high mass systems, however, placed planet e about four standard deviations away from each of its observed coordinates. Marois et al. interpreted these results to mean that the system is young and lower planetary masses are preferred. Recently, Currie et al. (2011) carried out similar integrations for a 10-13-13-13 $M_{Jup}$ configuration, corresponding to an age of 160 Myr. Their results also favor a young system with lower planetary masses. Other investigations have addressed the limits on the planetary masses and the age of the planetary system through dynamical simulations (Fabrycky & Murray-Clay 2010, Marshall et al. 2010, Goździewski &

Migaszewski 2009, Moyo-Martín et al. 2010, Reidemeister et al. 2009). However, the results of these investigations precede the discovery of the fourth planet.

In examining the results of the dynamical simulations reported by Marois et al. (2010) and Currie et al. (2011), we noticed the following issue with their four-planet systems. As reported by Marois et al. (2010), the inclination for planet e is 30° whereas the inclinations of the three-planet systems on which their four-planet systems are based are all 0°. These authors did not discuss the transformation of the three-planet systems to an inclination of 30° and did not report the orbital elements of planets b, c, and d. A system that is stable at one inclination may not be stable at a higher inclination if the system is required to remain consistent with the astrometric data.

The wide range of possibilities for the age and inclination of the planetary system orbiting HR 8799, and the release of additional astrometric data (Bergfors et al. 2011 and references therein) following the announcement of the discovery of the fourth planet in the system, prompted us to re-examine the stability of the planetary system in an attempt to further constrain its inclination and set a lower bound on the upper limit to the age of the system from a dynamical point of view. We explain our method for carrying out our dynamical simulations in Section 2, and, in Section 3, we present the results. We conclude this study in Section 4 by presenting a summary of our analysis and our final remarks.

## 2. METHOD

Our approach to constraining the age and inclination of the HR 8799 planetary system was to study its long-term stability by numerically integrating the orbits of the planets for different combinations of orbital elements. At this time, the astrometric data permit such a wide range of orbital elements that the free parameter space for each planet is too large to be adequately sampled in a reasonable period of time. Therefore, we made certain assumptions to reduce the range and number of orbital elements under consideration.

We assumed that the orbits of planets b and c are circular, a common assumption in the literature, and that their orbital elements can be specified with little error by their coordinates on 2008 August 12 as given by Marois et al. (2008). We chose 2008 August 12 for two reasons: simultaneous data for planets b, c, and d are available on this date, and the coordinates for these three planets on this date have the lowest uncertainties, corresponding to less than 0.3% in the semi-major axis of a circular orbit. In effect, with these assumptions, we held constant the orbital elements of planets b and c (for any given angle of inclination), thereby reducing the free parameter space to only the orbital elements for planets d and e. We further assumed the orbits of all four planets are coplanar, another common assumption in the literature, and the longitude of the ascending node for each orbit is 0°. We considered only the high-mass configuration for the planets (7, 10, 10, and 10 $M_{Jup}$), and we carried out integrations only for inclinations of 0°, 13°, 18°, 23°, and 30°.

At the start of our study (June 2011), simultaneous astrometric data for all four planets obtained with the same instrument were not available. (Galicher et al. 2011 have since published astrometric data for planets b, c, and d obtained on the same date and using the same instrument as observations of planet e reported by Marois et al. 2010.) We therefore adopted a two-stage approach in which we first determined the most stable three-planet system at each inclination then searched for the most stable four-planet system.

In the first stage, we considered only variations in the orbital elements for planet d and constructed a low-resolution grid of semimajor axis, eccentricity, longitude of pericenter, and mean anomaly. The semimajor axis varied depending on the inclination angle (see Figure 1) in increments of 0.1 AU. The eccentricity varied between 0 and 0.1 in increments of 0.01, and both the argument of pericenter and the mean anomaly varied between 0° and 360° in increments of 1°. To further reduce the number of orbital elements under consideration, we calculated the coordinates for planet d for every point in this grid and compared them to the observed coordinates on 2008 August 12 ($x = 8.51 \pm 0.08$ AU, $y = -22.93 \pm 0.08$ AU; Marois et al. 2008). Note that $+x$ is to the right (West on the sky) and $+y$ is up (North on the sky) in our coordinate system. If the coordinates for any grid point were more than one standard deviation outside the observed coordinates, we removed the grid point from consideration. This left us with approximately 2,400 combinations of orbital elements to consider at each angle of inclination.

We used the Bulirsch-Stoer algorithm in the *N*-body integration package MERCURY (Chambers 1999) to integrate each combination of orbital parameters (each point from the grid for planet d combined with fixed orbital elements for planets b and c as described above) for 200 Myr.

Although the high-mass configuration for the planets corresponds to an age of ~60 Myr in the hot-start planet cooling models, and therefore we could have stopped the integrations at 60 Myr, we chose 200 Myr as it represents a balance between being short enough to search the grid in a reasonable period of time and long enough to allow us to isolate the most stable three-planet systems. We assumed that the more stable the three-planet system, the more likely it will remain stable when we add a fourth planet. We held constant the mass of the star at 1.5 $M_{solar}$ (Gray & Kaye 1999) and the distance to the star at 39.4 pc (ESA 1997). We set the initial time step for all integrations to 200 days.

Next, we compared the orbits of the planets in the most stable three-planet systems to all of the astrometric data then available (June 2011), which did not include the data from Galicher et al. (2011) and Soummer et al. (2011). We identified the system at each angle of inclination most consistent with the astrometric data and ran the orbital elements in that system forward in time to 2009 August 1. We chose this date as the starting point for the second stage in our two-stage approach, because the observed coordinates on this date best represent the trend in all of the astrometric data for planet e. The uncertainties in the astrometric data for planet e are nearly identical, so no particular coordinate serves to further reduce the parameter space. We note that, when we add the astrometric data from Galicher et al. (2011) and Soummer et al. (2011) to the mix, the three-planet systems that were most consistent with the astrometric data during our study remain the most consistent of our systems with respect to all of the astrometric data currently available.

In the second stage in our two-stage approach, we added planet e to the most stable, three-planet system (determined as described above) at each angle of inclination. For planet e, we constructed a low-resolution grid of semimajor axis, eccentricity, argument of pericenter, and mean anomaly. Again, the semimajor axis varied depending on the inclination angle (see Figure 2) in increments of 0.1 AU. The eccentricity varied between 0 and 0.1 in increments of 0.01, and both the argument of pericenter and the mean anomaly varied between 0° and 360° in increments of 1°. We eliminated from consideration any point in the grid that placed planet e more than one standard deviation outside its observed coordinates on 2009 August 1 ($x = -11.9 \pm 0.5$, $y = -8.2 \pm 0.5$; Marois et al. 2010). This left us with approximately 160,000 grid points at each angle of inclination. This large number of grid points is a consequence of the higher uncertainties in the astrometric data. We carried out the *N*-body integrations in the same fashion as in the three-planet systems for all of the orbital elements for planet e in the grid and identified the most stable four-planet systems at each angle of inclination.

In the next section, we explain the results of our *N*-body integrations in each of the two stages in our method and discuss their implications for the age and inclination of the system.

## 3. RESULTS

*3.1 The Three-Planet Systems*

The results of the simulations of our three-planet systems appear in Figure 1. The *N*-body integrations returned 15 systems at an inclination of 0°, two at 13°, three at 23°, and four at 30° that remained stable for 200 Myr. No system remained stable for 200 Myr at 18°, but one system did remain stable for 160 Myr. As shown in Figure 1, at low angles of inclination, an island of stability appears at the lowest, initial semimajor axes, and at initial eccentricities between 0.04 and 0.08. As the inclination increases, this island of stability moves to higher semimajor axes, and slightly higher eccentricities.

Note that for each pairing of semimajor axis and eccentricity shown in Figure 1, many combinations of the argument of pericenter and mean anomaly were included in the grid (each angle varies from 0° to 359° in 1° increments). Typically, on the order of 10 combinations of the argument of pericenter and mean anomaly were consistent with the observed coordinates of planet d on 2008 August 12, and the sum of the angles remained within a few degrees of 290°. Each point in Figure 1, though, represents only one combination of semimajor axis, eccentricty, argument of pericenter, and mean anomaly – the combination for which the lifetime of the system is the longest.

For each of the 25 most stable, three-planet systems mentioned above, we checked the agreement between the coordinates of the planets in the system and the observed coordinates on the dates for which astrometric data were, at the time (June 2011), available. In other words, we performed a reduced $\chi^2$ test of the coordinates of the planets against the astrometric data. (We caution the reader to note that we did not attempt to model the astrometric data through a $\chi^2$ minimization procedure, which is a common practice in the literature.) We selected the system at each angle of inclination with the lowest reduced $\chi^2$ value to advance to the next stage in our two-stage approach. When compared to all of the astrometric data currently available, the systems that we selected during our study still yield the lowest reduced $\chi^2$ values: 1.1, 1.3, 1.3, 1.4, and 2.0 at 0°, 13°, 18°, 23°, and 30°, respectively. Since these three-planet systems are the foundation for our four-planet systems, we will discuss how the system coordinates compare to the observed coordinates in more detail in section 3.2.

Lastly, we note that we made no attempt to tweak these systems for a particular mean motion resonance. The most stable systems with the lowest reduced $\chi^2$ values at $i$=18°, 23°, and 30°, however, do exhibit an initial 1d:2c MMR to within 2%.

*3.2 The Four-Planet Systems*

The low-resolution, four-planet *N*-body integrations produced no system that remained stable for more than ~30 Myr. Plots of the lifetimes for the most stable systems for each pairing of the initial semimajor axis and eccentricity of planet e appear in Figure 2. As with the three-planet systems, many combinations of the argument of pericenter and mean anomaly were included in the grid, but only the combination with the longest lifetime is shown in Figure 2. For planet e, the sum of the two angles remained within a few degrees of 320°.

As shown in Figure 2, no region of the parameter space appears particularly stable at inclinations of $i$=0° and 13°. The longest lifetimes for systems at these inclinations were ~5 Myr, therefore, we did not conduct any further investigations in this region. At $i$=18°, 23°, and 30°, an island of stability appears near the center of each grid. At $i=30°$, a second island of stability emerges at higher eccentricities and lower semimajor axes.

To obtain more precise orbital parameters for planet e, we increased the resolution of our grid by an order of magnitude in the vicinity of each orbital configuration that had the longest lifetime within each island of stability in the low-resolution grid, and, again, carried out 200 Myr integrations. We repeated this process once more to achieve a precision of at least three significant digits in the orbital elements for planet e. (We refer to the final grid of orbital elements coming out of this process as the high-resolution grid.)

Figure 3 shows the lifetimes for the most stable systems for each pairing of initial semimajor axis and eccentricity in the high-resolution grid mentioned above. At inclinations of $i=18°$ and $i=23°$, the longest lifetimes are ~41 Myr and ~46 Myr, respectively. At $i=30°$, the longest lifetimes are ~31 Myr and ~155 Myr, with the lower value corresponding to the central island in Figure 2. We note that a total of 17 systems at $i=30°$ had lifetimes in excess of 60 Myr, the lower bound for the age of the planetary system in the high-mass configuration. The orbital elements of these systems are all within a few percent of one another (see Figure 3), so we consider only the system with the longest lifetime, 155 Myr, forthwith. The orbital elements for the most stable four-planet systems within each of the four islands of stability appear in Table 1. We note that the semimajor axes for planet e for the two systems at an inclination of 30° appear, in the first case, much larger, and, in the second case, much smaller than the observed separation between planet e and HR 8799. In the first of these two systems, planet e is near periastron, and, in the second, planet e is near apastron. To portray the (in)stability of the system with the 155 Myr lifetime, we prepared a time-series plot of the semimajor axes of the planets, shown in Figure 4. The semimajor axes of the orbits oscillate with high amplitude but remain stable up until the last fraction of a percent of the lifetime of the system. Here, as in the other three systems, instability sets in quite suddenly.

Reduced $\chi^2$ tests of the coordinates of the planets in each system compared to all of the astrometric data currently available yield values that range between 1.2 and 1.9 (see Table 1). Figure 5 compares the coordinates of the planets in each system on the dates for which astrometric data are currently available to the observed coordinates. Two trends are immediately apparent: (1) In the most-stable four-planet systems that we obtained through our numerical integrations, the higher the inclination, the more the coordinates for planets b and c diverge from the coordinates determined from archival *Hubble Space Telescope* data by Soummer et al. (2011), marked "1998" in Figure 5. These archival data were released after we completed our integrations. In the worst case, at $i=30°$, the coordinates of planets b and c in our systems are three to four standard deviations away from the observed coordinates. This suggests that either these planets are in eccentric orbits or their orbital inclinations are lower than that of planet d. (2) The orbits for planet e in our most stable systems place the planet closer to the star than is observed. Moreover, the more stable the system, the greater the separation of planet e from the other planets.

## 4. CONCLUSIONS

We conducted a dynamical study of the four planets orbiting HR 8799 in the high-mass configuration (7-10-10-10 $M_{Jup}$) in an attempt to constrain the inclination and the age of this system. We carried out more than 1.5 million numerical integrations of different combinations of the orbital elements of the planets, and identified the most stable four-planet systems at inclinations of $i$=18°, 23°, and 30°. In general, the system lifetimes did not exceed 30 to 45 Myr, suggesting that the planets have lower masses and the system is young. We found, however, a number of high-mass systems, all relatively close to one another in parameter space, at $i$=30° having lifetimes in excess of 60 Myr, which is the expected age of the system in the high-mass configuration. The most stable of these systems has a lifetime of ~155 Myr.

Reduced $\chi^2$ tests, comparing the coordinates of the planets in each of the four most stable systems to the astrometric data, yield values greater than one. This is not unexpected because of the scatter in the astrometric data and the potential for systematic differences between the astrometric data obtained with different instruments.

The results of our simulations point to an interesting trend in the coordinates of planet e compared to the observed coordinates. The longer the lifetime of the system, the greater the difference between these coordinates in the direction of the star. This suggests, as expected, that the planets must have lower masses in order to remain in such close proximity for longer periods of time. Furthermore, the orbits of planets b and c in the higher inclination systems diverge significantly from recently released archival data (Soummer et al. 2011), which suggests that the assumptions of circular orbits for planets b and c and coplanar orbits for all of the planets must be relaxed in order to portray a more accurate picture of the orbital architecture of this system. Of course, relaxing these assumptions dramatically increases the free parameter space, therefore requiring a much lengthier search for stable systems.

In closing, we note that, although our approach has the advantage of dramatically reducing the parameter space with little computational overhead, it has some distinct disadvantages as well. Having either fixed or constrained the orbital parameters for each planet using a single astrometric data point, we have excluded regions of the parameter space consistent with the astrometric data on the whole. However, such measures are necessary in order to test different orbital architectures for the system in a reasonable period of time.

This project has been partially funded by a Support and Development Award from the West Chester University College of Arts and Sciences and by the West Chester University Department of Physics to JJS. NH acknowledges support from the NASA Astrobiology Institute under Cooperative Agreement NNA09DA77 at the Institute for Astronomy, University of Hawaii, and NASA EXOB grant NNX09AN05G. We thank J.T. Singh, Coordinator of Technical Support Services in Academic Computing at West Chester University, for contributing decommissioned computers to this project. We also thank the following undergraduate students at West Chester University who helped us to complete the construction of the computer network used in this project and to conduct some early investigations of the HR 8799 system: Steve Assalita, Brittany Johnstone, Nora Pearse, Michael Savoy, and Michael Scott.

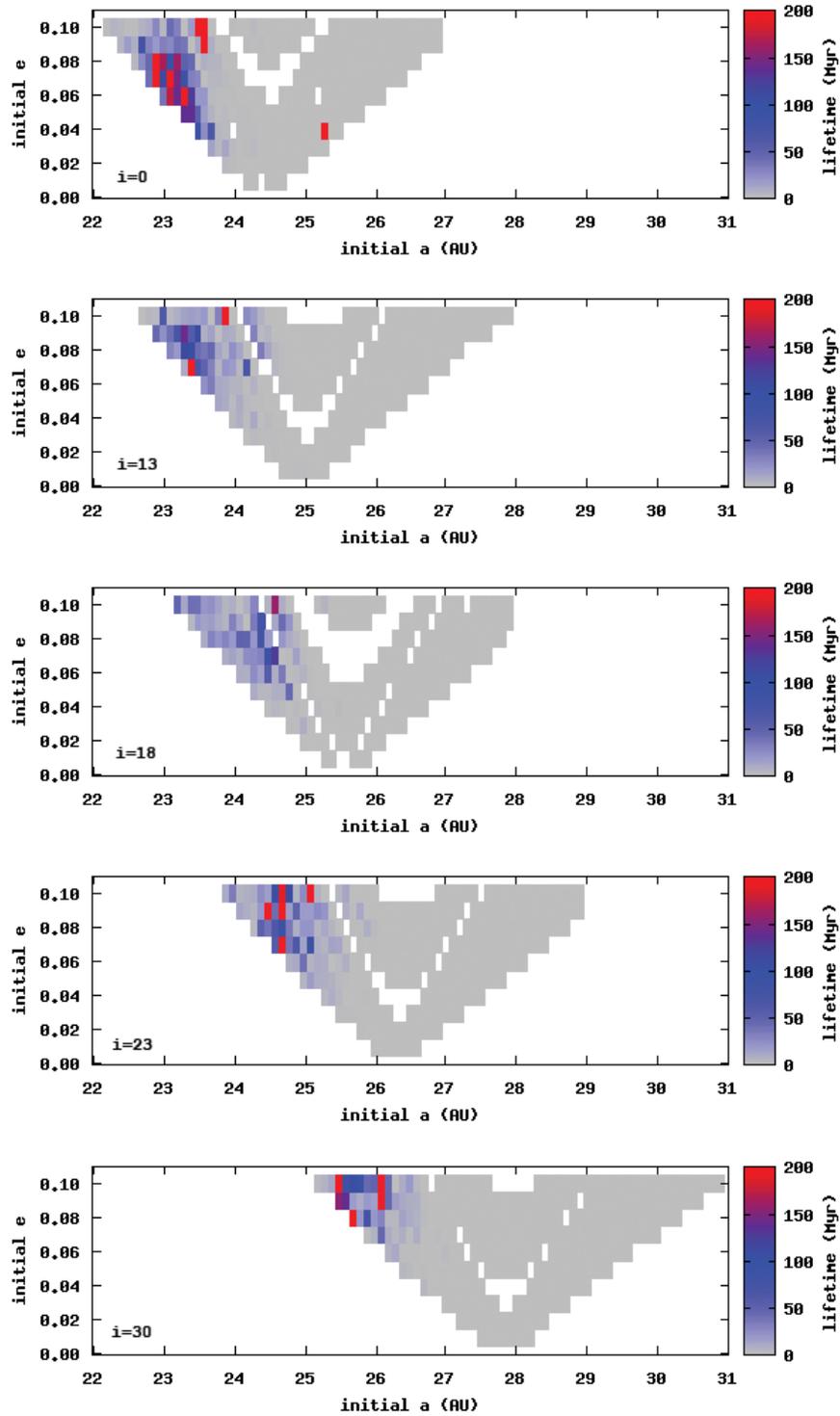

**Figure 1.** Longest lifetimes for each pairing of the initial semimajor axis and eccentricity of planet d in the low-resolution grids at all inclinations. The missing data in the characteristic "V" shape in each plot represent occasions when the integrator stopped before concluding the integration due to multiple close encounters between planets. Such systems are not stable, so we made no attempt to recover the data.

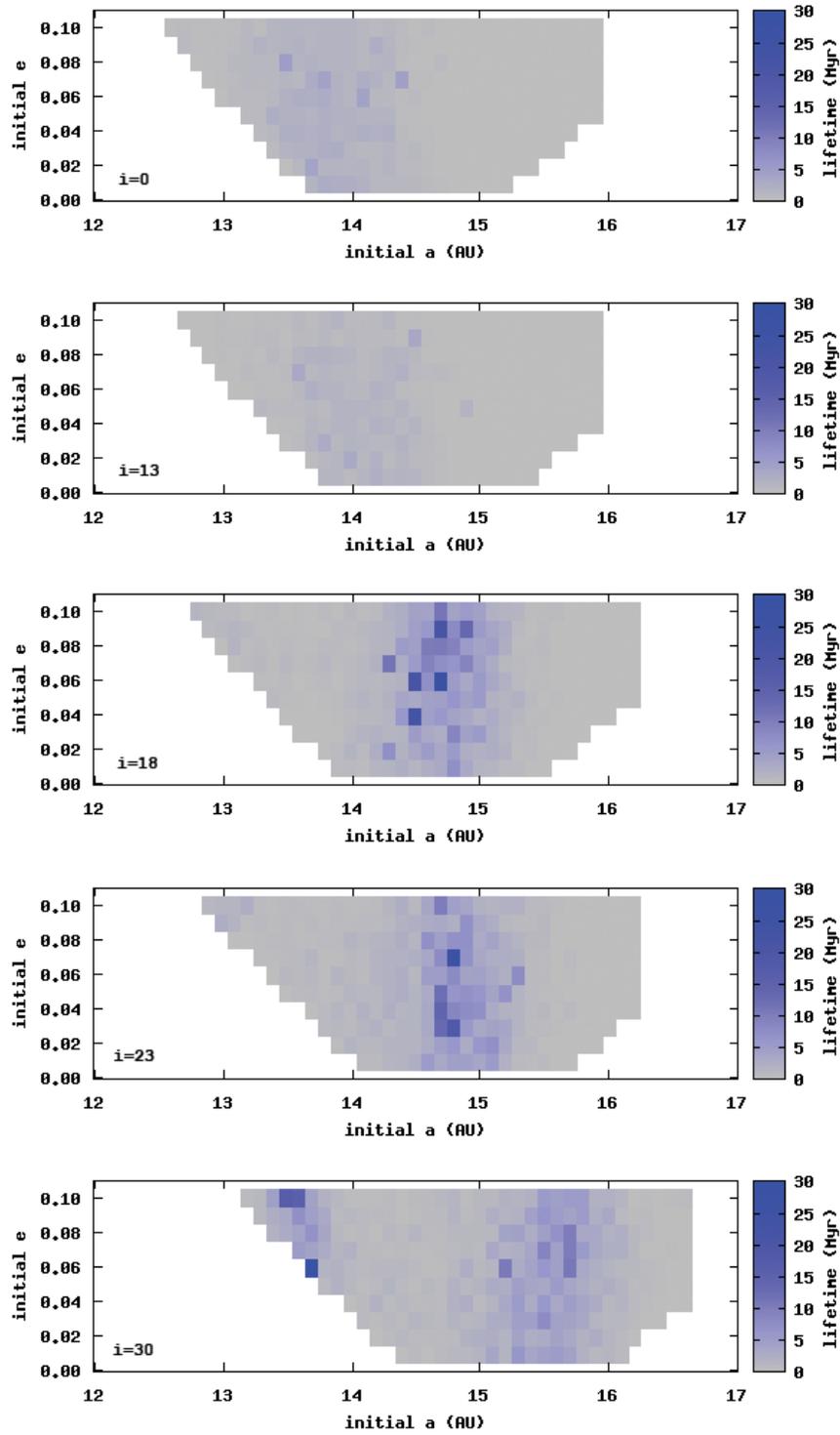

**Figure 2.** Longest lifetimes for each paring of the initial semimajor axis and eccentricity of planet e in the low-resolution grids at all inclinations.

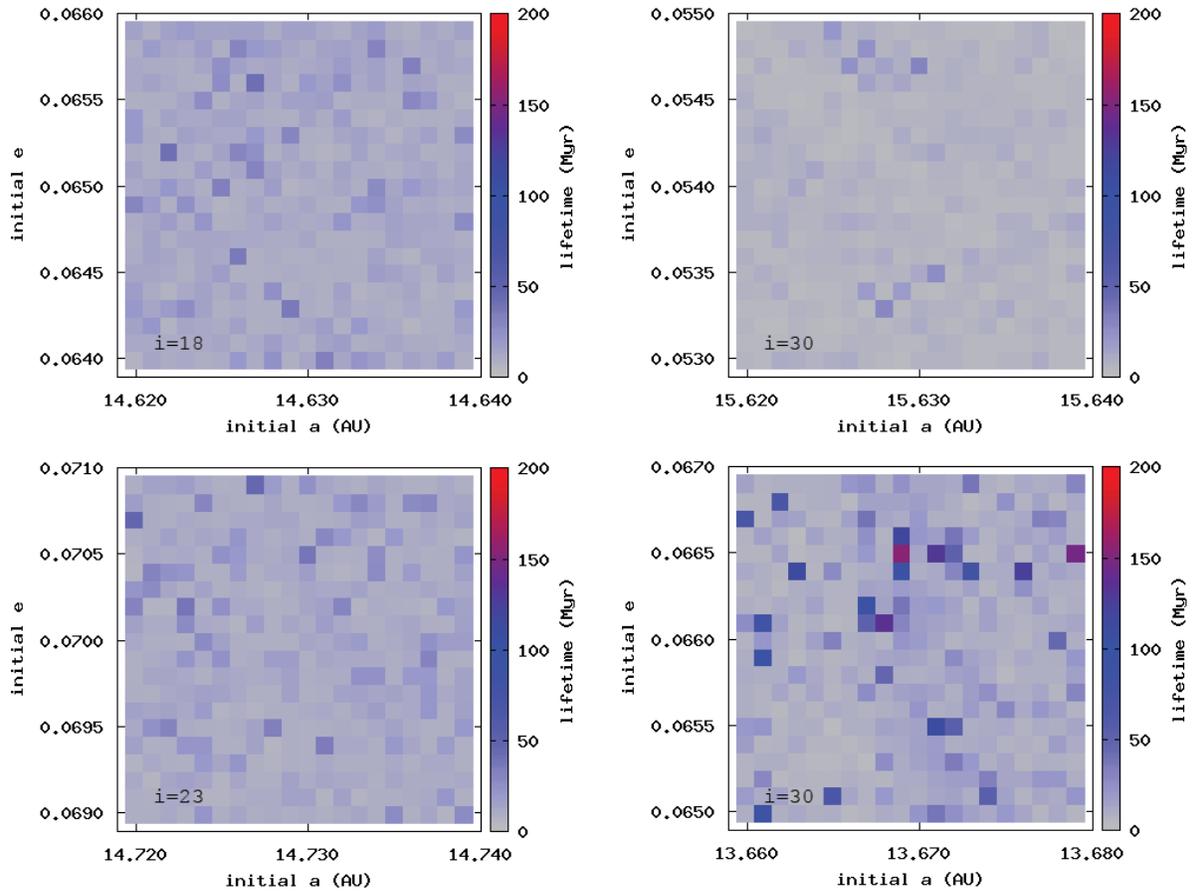

**Figure 3.** Longest lifetimes for each pairing of the initial semimajor axis and eccentricity of planet e in the high-resolution grids at *i*=18°, 23°, and 30°.

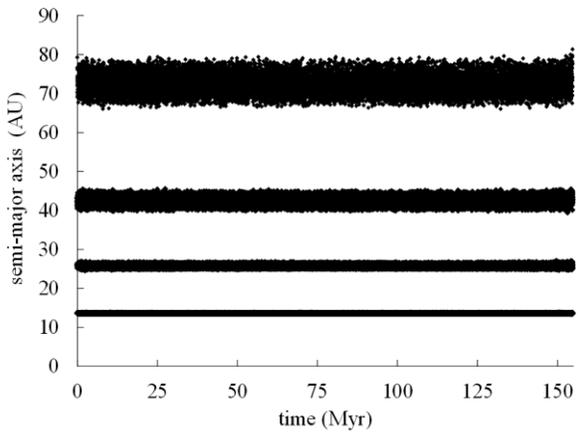

**Figure 4.** Semimajor axis of planet e as a function of time for the 155 Myr system.

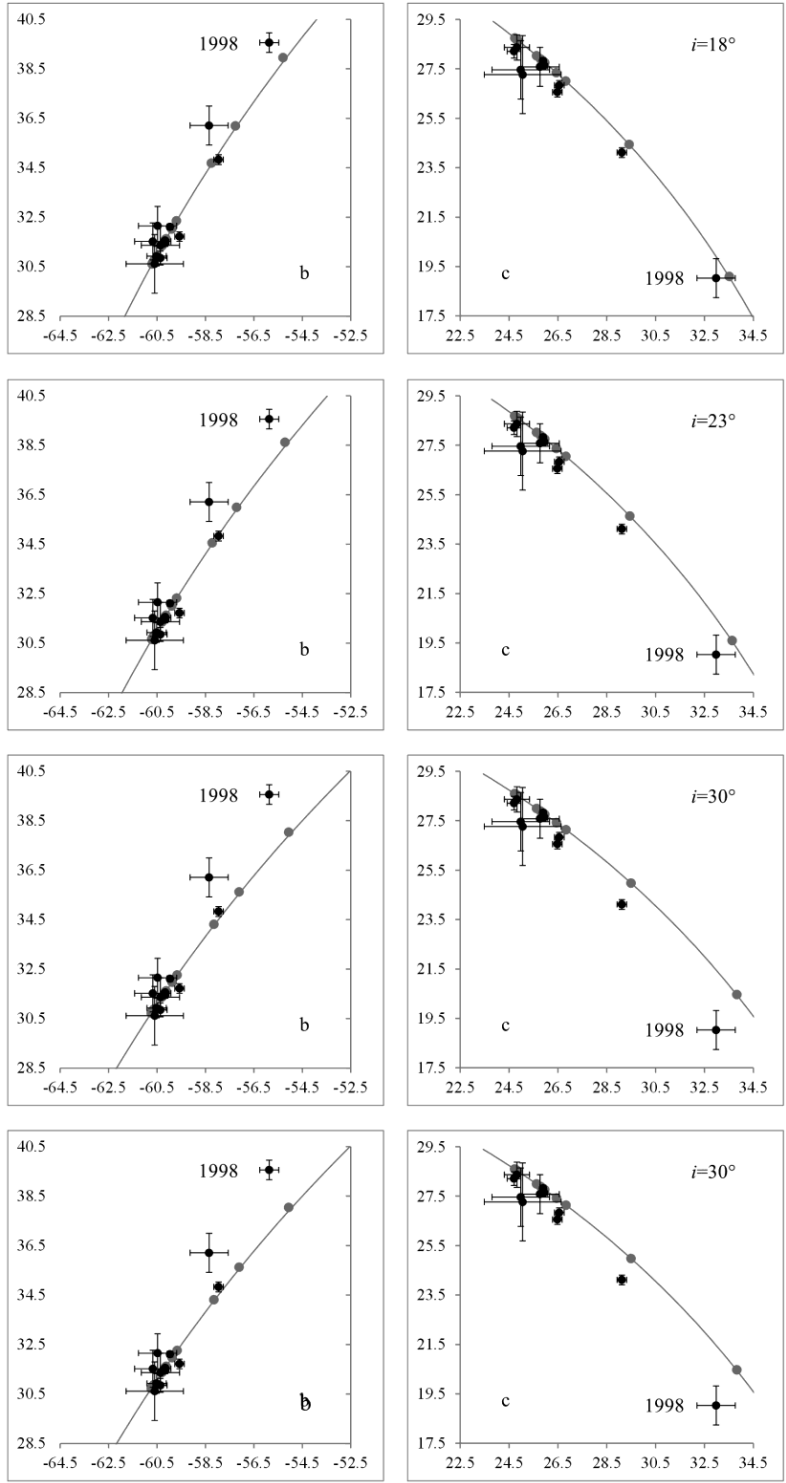

**Figure 5 (left).** Coordinates of planets b, c, d, and e in the four most stable integrated systems (gray circles) compared to their observed coordinates (black circles). The order of the systems here is the same as it is in Table 1 from top to bottom. The gray line in each plot represents the initial orbit of the planet.

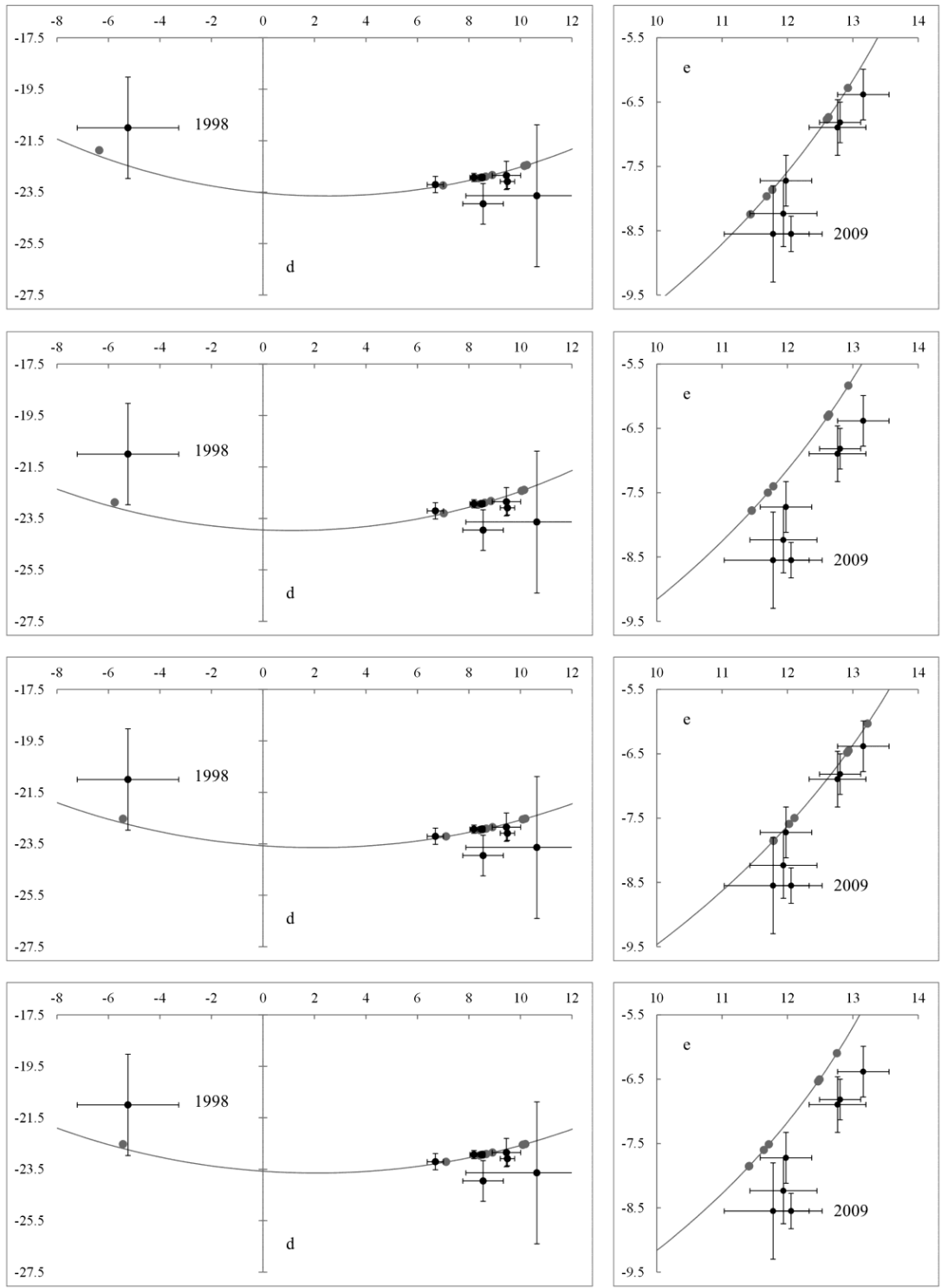

**Figure 5 (right).** The coordinates of the planets in the integrated systems diverge from their initial orbits due to interactions between the planets. Note that $+x$ is to the right (West on the sky) and $+y$ is up (North on the sky) in our coordinate system.

**Table 1.** Orbital elements[1] (epoch 2009 August 1) for the high resolution, four-planet systems having the longest lifetimes. The quantity $\tilde{\chi}^2$ represents the value of the reduced chi-squared test of the coordinates of the system compared to the observed coordinates. The quantity $L$ represents the lifetime of the system.

| $i$ (°) | planet | $a$ (AU) | $e$ | $\omega$ (°) | $M$ (°) | $\tilde{\chi}^2$ | $L$ (Myr) |
|---|---|---|---|---|---|---|---|
| 18 | b | 66.6877 | 0.000508 | 25.3977 | 126.4287 | 1.2 | 41 |
|  | c | 39.0838 | 0.000681 | 39.3724 | 10.8584 |  |  |
|  | d | 24.5965 | 0.100090 | 170.9291 | 111.5816 |  |  |
|  | e | 14.627 | 0.0656 | 245.96 | 69.65 |  |  |
| 23 | b | 69.2278 | 0.000471 | 24.8001 | 126.2229 | 1.5 | 46 |
|  | c | 39.8149 | 0.000642 | 39.2822 | 11.8263 |  |  |
|  | d | 24.6966 | 0.070147 | 127.9827 | 161.5065 |  |  |
|  | e | 14.720 | 0.0707 | 257.78 | 58.59 |  |  |
| 30 | b | 70.3239 | 0.000398 | 24.6530 | 124.8323 | 1.8 | 31 |
|  | c | 41.2749 | 0.000568 | 39.3619 | 13.3745 |  |  |
|  | d | 26.0967 | 0.090127 | 146.9603 | 137.2508 |  |  |
|  | e | 15.630 | 0.0547 | 293.44 | 26.07 |  |  |
| 30 | b | 70.3239 | 0.000398 | 24.6530 | 124.8323 | 1.9 | 155 |
|  | c | 41.2749 | 0.000568 | 39.3619 | 13.3745 |  |  |
|  | d | 26.0967 | 0.090127 | 146.9603 | 137.2508 |  |  |
|  | e | 13.669 | 0.0665 | 144.25 | 176.92 |  |  |

[1] The longitude of the ascending node is 0° in all cases. Note that $+x$ is to the right on the sky (West) and $+y$ is up (North) in our coordinate system.